\documentstyle[12pt,preprint]{aastex}
\begin{document}

\title{Plasma Physics Processes of the Interstellar Medium}

\author{Steven Spangler (University of Iowa), Marijke Haverkorn (Netherlands Foundation for Research in Astronomy), Thomas Intrator (Los Alamos National Laboratory), Russell Kulsrud (Princeton University), Alex Lazarian (University of Wisconsin), Seth Redfield (Wesleyan University), and Ellen Zweibel (University of Wisconsin)}

\section{Introduction} This ``white paper'' is directed to the GAN (galactic neighborhood) theme of the Decadal Survey.  We are particularly interested in interstellar medium (ISM) phenomena which can be described by the laws of plasma physics.  This is appropriate since nearly all of the ISM phases possess sufficient ionization for plasma behavior. Our advocacy of this position comes from the conviction that many of the important processes and phenomena which occur in the ISM may elude satisfactory understanding until the plasma physics aspects are recognized, and results and methods of plasma physics brought to bear on them. Study of the interstellar medium is a major disciplinary area of astronomy and astrophysics.  It is here where the process of star formation occurs.  The cosmic rays are probably accelerated in the ISM, and its properties certainly govern their propagation and diffusion. In the rest of this document, we identify some of the most interesting and important topics in the astrophysics of the interstellar medium, which merit attention by the astronomy community in the next decade.  {\em The theme of this white paper was chosen by the Topical Group on Plasma Astrophysics of the American Physical Society.}

We state at the outset that a major recommendation of this white paper is that the astronomy and astrophysics community very seriously consider supporting plasma laboratory work.  In the past ten years, laboratories have been built which address important fundamental physical processes such as magnetic reconnection, nonlinear interaction of Alfv\'{e}n waves and turbulence, magnetohydrodynamic instabilities, and the dynamo generation of magnetic fields.  The information coming from these experiments can illuminate important astronomical phenomena.  It is entirely possible that the most important discovery in astronomy of the next decade will not come from an astronomical telescope, but from a physics laboratory experiment. The last section of this white paper lists some current laboratory experiments which investigate astronomically-relevant problems. 
\section{Outstanding Issues in the Physics of the Interstellar Medium}
\subsection{Properties of the Interstellar Magnetic Field}
Interstellar magnetic fields play a pivotal role in many physical processes in the interstellar medium such as cosmic ray acceleration and propagation, the ISM energy balance, and star formation \citep{Ferriere01}. Despite their importance, relatively little is known about the strength and structure of these magnetic fields in the Milky Way because the observations are complex and necessarily indirect.

One of the more fruitful diagnostics of cosmic magnetic fields has been analysis of polarized radio emission. This emission comes from extragalactic (radio galaxies and quasars) and Galactic sources (pulsars and diffuse synchrotron emission from cosmic ray electrons). The emission is subject to Faraday rotation and is partially depolarized during propagation through the Galaxy. This rotation and depolarization yields information on the strength and structure of the Galactic field. 

The coming years will be a golden era for interstellar magnetism studies, due to recent advances in radioastronomical techniques such as rotation measure synthesis \citep{Brentjens05} as well as the prospect of major new radio telescopes which are being planned or are under construction. Interstellar magnetism is an important science driver for these telescopes. For example, ``Cosmic Magnetism'' has been named a Key Science Project for both for the Square Kilometer Array (SKA) and the LOw Frequency ARray (LOFAR).
Development of the SKA and its pathfinders (LOFAR, MWA, etc), particularly their polarization properties, is of foremost importance for the study of cosmic magnetism.
\subsection{Structuring of the Interstellar Medium by Winds and Shocks}
There is abundant observational evidence that stellar winds can produce cavities in the interstellar medium.
On larger scales, \cite{Everett08} have shown that steady-state galactic winds can be driven from the inner Milky Way by a combination of thermal and cosmic ray pressures.  In this case, the 
wind should draw up the interstellar magnetic
field lines, creating apparent ``holes'' in the 
horizontal component of the galactic magnetic
field.  Such holes could provide (a) a possible escape route for the 
 cosmic rays, and (b) an impact
on galactic dynamos.  Whether this occurs
in our Galaxy or not seems uncertain, and depends on the ability of stellar winds and supernovae to keep a ``galactic fountain'' going \citep{MacLow99,deAvillez01,Joung06}.
A galactic
wind would correspond to a permanent hole in the 
horizontal field, unless it is closed by magnetic reconnection, which
would take a long time. 

Gaining certain knowledge of these flows is
really essential for understanding galactic cosmic rays
and the galactic dynamo.  The fountain would lead
to the low mean density derived for the medium
through which cosmic rays propagate, as found from
radioisotope measurements.  It would also help answer the
question of whether the propagation of cosmic rays 
is one or three-dimensional.  The boundary condition
on the galactic dynamo tacitly assumes the escape
of net flux out of the galaxy,  without which
the dynamo cannot work.  

Progress on these issues has been gained through observations of the
high latitude background H alpha emission \citep[e.g.][]{Levine06}.  
These results may lead to a better understanding of 
this expulsion of the interstellar medium into
the halo or the extragalactic medium.  
It would seem that a combined effort by
plasma astrophysicists and observational astronomers 
could make a very significant impact on these
questions in the next ten years.
\subsection{Turbulence in the Interstellar Medium}
The interstellar medium is a partially-to-fully ionized gas (the degree depending on the phase under consideration) with embedded sources of energy and an overall velocity shear due to differential galactic rotation.  As such, turbulence is expected, and indeed there is much observational evidence for its existence \citep[e.g.][]{Franco99}. 

This turbulence almost certainly plays an important role in the astrophysics of the ISM.  Turbulence-associated magnetic field fluctuations govern the diffusion and propagation of charged particles in the galaxy.  Turbulence dissipates, and yields its energy to its host fluid as heat.   In the ISM, the turbulence dissipation timescale can be much shorter than other timescales, making turbulent dissipation an effective heating mechanism\footnote{Obviously the mechanism generating the turbulence must also be identified.  Turbulence could be viewed as a catalyst, effectively transferring energy from some cosmic free energy source to the thermal heat of the host medium.}.  Finally, as is the case in the solar wind and corona, where the observations are better than those of the ISM, interstellar turbulence is almost certainly responsible for determining fundamental transport coefficients such as viscosity, resistivity, and thermal conductivity. Knowledge of the values of these transport coefficients is crucial for developing credible mathematical models of the ISM.  
There are many types of radio and optical observations which help us diagnose interstellar turbulence, such as radio scintillations and the analysis of spectral data cubes of 21 cm HI observations \citep{Lazarian08}.  

Progress in this field will be facilitated by radio observations with higher sensitivity (provided by instruments such as the SKA), and with moderate sensitivity and modern receiver and control electronics at low frequencies, such as the MWA (Murchison Widefield Array) and LWA (Long Wavelength Array).  These instruments will permit scintillation observations of pulsars which have been too faint for this type of measurement, at least in the low frequency regime.   
\subsection{The Plasma State of the Interstellar Medium}
To understand the evolutionary processes which govern the interstellar medium, it is necessary to know its basic physical properties, and the variation in these properties from one place to another.  These basic properties consist of the fundamental ``plasma parameters'' such as electron density, elemental composition, ionization fraction, magnetic field strength, electron temperature, ion temperature, and (if possible) the functional form of the electron and ion distribution functions.  

The two best-diagnosed parts of the ISM are the Diffuse Ionized Gas, a tenuous gas that fills perhaps 25 \% of interstellar space, and the local interstellar medium.
The local interstellar medium (LISM) is the interstellar material that
resides in close ($\sim$100\,pc) proximity to the Sun.  Proximity
is a special characteristic that drives much of the interest in the
LISM.  First, proximity provides an opportunity to observe general ISM
phenomena in great detail, and in three dimensions.  Knowledge of general ISM phenomena
in our local corner of the galaxy, can be applied to more distant and
difficult to observe parts of the universe.  Second, proximity implies
an interconnectedness.  For example, this relationship between the Sun
(and planets), and our surrounding interstellar environment is
evidenced by the structure of the heliosphere, the cosmic ray flux in
the inner solar system, etc.

High resolution spectroscopy ($R \sim $100,000 -- 300,000) in the
optical and ultraviolet is absolutely critical to advancing on the promising results 
of the past several years.  This
instrumentation also needs to be placed on large aperture telescopes.
Due to the close proximity of sight lines, faint background stars must
be used, and high signal to noise is required to adequately model the
absorption profiles.  A large number of sight lines are required,
(i.e., a homogeneous, large-scale survey), in order to probe the wide
range of physical scales that we know contribute to the structure of
the ISM, from $< 1$ AU to tens of parsecs.

\subsection{Magnetic Field Reconnection in the Interstellar Medium}
Magnetic field reconnection is one of the foremost topics in basic plasma physics and astrophysical plasma physics. The importance of reconnection in astrophysics is that the considerable energy in magnetic fields can be released, plasma flows generated, and perhaps particles accelerated in electric fields generated in the reconnection regions.  It is clear that magnetic reconnection is occurring and is crucial to phenomena on the Sun, and is involved in the evolution of active regions, solar flares, and coronal mass ejections.  

It seems that magnetic reconnection and its associated current sheets must arise in the interstellar medium. Faraday rotation measurements of extragalactic radio sources show random variations from one line of sight to another \citep[e.g.][]{Haverkorn04}, and these variations are generally interpreted as evidence for a random magnetic field which is of order or larger than the systematic one.  Caution is necessary in this interpretation, because Faraday rotation is determined by the electron density as well as the magnetic field, and the relative contribution of electron density and magnetic field fluctuations is not completely specified.  This topic itself should attract attention in the next decade.  However, if the Faraday rotation fluctuations do indeed indicate a highly stochastic field, stochastic current sheets should fill the ISM.    

The existence and consequences of this interstellar reconnection should be one of the main themes of the astrophysics of the interstellar medium during the next decade.  One simple way in which this goal could be advanced is simply by a greater awareness of work in the fields of basic plasma physics, solar physics, and heliospheric physics, bearing in mind the different parameter regimes of the ISM.  Observational research  could consist of additional, higher spatial-density Faraday rotation measurements of extragalactic radio sources to identify regions of high magnetic shear \citep[e.g.][]{Brown01}.  These regions could then be inspected for other anomalies, such as enhanced heating or plasma flows.  These observations could be made with the upgraded Very Large Array (EVLA), and eventually with the Square Kilometer Array (SKA). Theoretical work can address the basic physical description of reconnection in the ISM, such as whether it can be properly described as collisionless or collisional, and also should make a connection with the large community working on magnetospheric, heliospheric, and solar reconnection \citep{Yamada07}.  Additional theoretical work can also illuminate the locations and circumstances under which reconnection occurs.  MHD simulations of the interstellar medium \citep[and others]{Vazquez03, MacLow03} show substantial shearing of the ISM field on large spatial scales.  These investigations cannot describe the scales or physical processes that would be involved in interstellar reconnection, but they are at least suggestive that reconnection current sheets may arise.  Finally, there are reasons to believe that interstellar turbulence (Section 2.3) might enhance the rate of reconnection in the ISM \citep{Lazarian99}.    
 \section{The Interstellar Medium in a Can: Laboratory Experiments of Relevance to ISM Astrophysics}
In the past 10 to 15 years, there have been significant advances in experimental plasma physics, which directly address the themes of ISM astrophysics discussed above. These advances have resulted from larger machines capable of including more relevant spatial scales \citep{Carter06}, as well as the introduction of more sophisticated diagnostic techniques, such as laser-induced fluorescence \citep{Skiff07}.  Spectroscopic techniques very similar to those employed in observational astronomy have also been used in the laboratory \citep{Cothran06,Denning08}.  There are many current experiments which are yielding results of relevance to ISM astrophysics, of which we note only a few.  The basic physics of reconnection is being studied in the MRX (Magnetic Reconnection Experiment, \cite{Yamada97}), SSX (Swarthmore Spheromak Experiment, \cite{Brown99}), and RSX (Reconnection Scaling Experiment, \cite{Furno03}) devices.  The nonlinear interaction of Alfv\'{e}n waves, a fundamental process in many areas of astrophysics, has been studied by \cite{Carter06} in the LAPD machine at UCLA, a device large enough to accommodate Alfv\'{e}nic physics. Flows in a laboratory device have been generated by magnetic effects \citep{Furno06}.  As a last example, the generation of magnetic fields from a non-magnetic, rotating fluid via a dynamo effect is now accessible to experiments in the Madison Dynamo experiment at the University of Wisconsin \citep{Spence06}.  These experiments can be substantially advanced in the next decade, and the astronomical community should enthusiastically support these projects . 


\begin{thebibliography}{}
\bibitem[Brentjens and de Bruyn (2005)]{Brentjens05} Brentjens, M.A. and de Bruyn, A.G. 2005, \aap~441, 1217
\bibitem[Brown (1999)]{Brown99} Brown, M.R.  1999, Phys. Plasm. 6, 1717
\bibitem[Brown and Taylor (2001)]{Brown01} Brown, J.C. and Taylor, A.R. 2001, \apj~563, L31
\bibitem[Carter et al (2006)]{Carter06} Carter, T.A., Brugman, B., Pribyl, P., and Lybarger, W. 2006, \prl~96, 155001
\bibitem[Cothran et al (2006)]{Cothran06} Cothran, C.D., Fung, J., Brown, M.R., and Schaffer, M.J. 2006, Rev. Sci. Inst. 77, 063504
\bibitem[de Avillez and MacLow (2001)]{deAvillez01} de Avillez, M.A. and MacLow, M.M. 2001, \apj~551, L57
\bibitem[Denning et al (2008)]{Denning08} Denning, C.M., Wiebold, M., and Scharer, J.E. 2008, Phys. Plasm. 15, 072115
\bibitem[Everett et al (2008)]{Everett08} Everett, J.E., Zweibel, E.G., Benjamin, R.A., McCammon, D., 
Rocks, L., and Gallagher, J.S. 2008 \apj~674, 258
\bibitem[Ferriere (2001)]{Ferriere01} Ferriere, K.M. 2001, Rev. Mod. Phys. 73, 1031
\bibitem[Franco and Carrami\~{n}ana (1999)]{Franco99} Franco, J. and Carrami\~{n}ana, A. 1999, Interstellar Turbulence, Cambridge University Press
\bibitem[Furno et al (2003)]{Furno03} Furno, I., and 9 authors 2003, Rev. Sci. Inst. 74, 2324
\bibitem[Furno et al (2006)]{Furno06} Furno, I., and 6 authors 2006, \prl~97, 5002
\bibitem[Haverkorn et al (2004)]{Haverkorn04} Haverkorn, M., Katgert, P., and de Bruyn, A.G. 2004, \aap~427, 169
\bibitem[Joung and MacLow (2006)]{Joung06} Joung, M.K.R. and MacLow, M.M. 2006, \apj~653, 1266
\bibitem[Lazarian and Vishniac (1999)]{Lazarian99} Lazarian, A. and Vishniac E. 1999, \apj~517, 700
\bibitem[Lazarian (2008)]{Lazarian08} Lazarian, A. 2008, \ssr (``online first''; tmp..195L)
\bibitem[Levine et al (2006)]{Levine06} Levine, E.S., Blitz, L., and Heiles, C. 2006, \apj~643, 881
\bibitem[MacLow and Ferrara (1999)]{MacLow99} MacLow, M.M. and Ferrara, A. 1999, \apj~513, 142
\bibitem[MacLow (2003)]{MacLow03} MacLow, M.M. 2003, LNP 614, 182
\bibitem[Redfield and Linsky (2004)]{Redfield04} Redfield, S. and Linsky, J.L. 2004, \apj~613, 1004
\bibitem[Skiff et al (2007)]{Skiff07} Skiff, F., Uzun, I., and Diallo, A. 2007, Plasm. Phys. Con. Fusion 49, B259
\bibitem[Spence et al (2006)]{Spence06} Spence, E.J., Nornberg, M.D., Jacobsen, C.M., Kendrick, R.D., and Forest, C.B. 2006, \prl~96, 055002
\bibitem[Vazquez-Semadini et al (2003)]{Vazquez03} Vazquez-Semadini, E., Gazol, A., Passot, T., and Sanchez-Salcedo, J. 2003, LNP 614, 182
\bibitem[Yamada et al (1997)]{Yamada97} Yamada, M. and 8 other authors, 1997, Phys. Plasm. 4, 1936
\bibitem[Yamada (2007)]{Yamada07} Yamada, M. 2007, Phys. Plasm. 14, 058102

\end{thebibliography}
\end{document}